# Fuzzy cellular model for on-line traffic simulation

Bartłomiej Płaczek

Faculty of Transport, Silesian University of Technology,
ul. Krasinskiego 8, 40-019 Katowice, Poland
`bartlomiej.placzek@polsl.pl`

**Abstract.** This paper introduces a fuzzy cellular model of road traffic that was intended for on-line applications in traffic control. The presented model uses fuzzy sets theory to deal with uncertainty of both input data and simulation results. Vehicles are modelled individually, thus various classes of them can be taken into consideration. In the proposed approach, all parameters of vehicles are described by means of fuzzy numbers. The model was implemented in a simulation of vehicles queue discharge process. Changes of the queue length were analysed in this experiment and compared to the results of NaSch cellular automata model.

## 1 Introduction

Road traffic models are often used in on-line mode to determine current and actual traffic parameters as well as to forecast future state of the flow for traffic control purposes. The term on-line means that the traffic model works in real time and gathers current traffic data acquired by vehicle detectors. An example can be given here of an on-line traffic model that uses counts of detected vehicles to calculate discharge time of a vehicles queue in a crossroad approach. The anticipation ability of traffic parameters is essential for actuated and synchronized traffic control.

Available systems of adaptive traffic control (e.g. SCOOT [1], UTOPIA [2]) use traffic models that describe queues or groups of vehicles rather than individual cars and their parameters. However, individual characteristics related to class of a vehicle are very important from the traffic control point of view, as they have a significant influence on traffic conditions and capacity of road infrastructure.

Modern traffic control systems are mostly intended for cooperation with traffic detectors that recognise presence or passing of vehicles and count them (e.g. inductive loops). Additional functionalities offered by vision based sensors cannot be fully utilised in systems of this kind [3]. Video-detection technology is usually simply adopted as a substitute for inductive loops, thus important data available for vision based sensors is discarded.

Properties of particular vehicles are considered when using microscopic traffic models. Computationally efficient and sufficiently accurate models have been developed based on the cellular automata theory [4]. In the literature there is a lack of information about experiments on cellular models implementation in traffic control systems.

In this paper a fuzzy cellular traffic model is introduced. This model was intended for on-line applications in traffic control. It enables utilisation of complex traffic data registered by vision based sensors. Development of this model was motivated by the following requirements.: (1) A vehicle influence on traffic conditions depends on its individual features, thus various classes of vehicles have to be modelled. (2) The traffic model has to provide data interfaces for many sources - detectors of different types. (3) The uncertainty has to be described in traffic model to take into account random nature of traffic processes as well as rough character of vehicles recognition (detection) results. (4) Computational complexity of the model has to be appropriately low to allow on-line processing.

The rest of this paper is organised as follows. Section 2 includes a brief survey of cellular automata applications in road traffic modelling. A fuzzy cellular traffic model is introduced in section 3. In section 4 simulation results of vehicles queue discharge process are discussed for the proposed model and compared with those of NaSch cellular automaton. Finally, in section 5 conclusions are drawn.

## 2   Cellular automata in traffic modelling

Due to their simplicity, cellular automata have become a frequently used tool for microscopic modelling of road traffic processes. Traffic models using cellular automata have high computational efficiency and allow sufficiently accurate simulation of real traffic phenomena [5].

The cellular automaton model for road traffic simulation was introduced by Nagel and Schreckenberg (NaSch) [4] in 1992. In this model the traffic lane was divided into cells of 7,5 m. Each vehicle is characterised by its position (cell number) and velocity (a positive discrete value lower than fixed maximum). The velocity is expressed as a number of cells that vehicle advances in one time step. Movement of vehicles in this model is described by a simple rule that is executed in parallel for each vehicle. It has the capability of mapping of real traffic streams parameters (fundamental diagram) and enables simulation of phenomena that can be observed in reality (e.g. traffic jam formation).

In the literature many traffic models can be found that are based on the Nagel-Schreckenberg concept. Numerous models have been introduced that uses so-called slow-to-start rules to reflect metastable state of traffic flow [6,7,8,9]. According to the slow-to-start (s2s) rule stopped vehicles accelerate with lower probability than moving ones. Different rules of this kind takes into account various factors: number of free cells in front of a vehicle (gap), present or previous state of a vehicle (velocity). In [10] a new set of rules has been proposed to better capture driver reactions to traffic that are intended to preserve safety on the highway.

On the basis of the NaSch cellular automaton multi-lane traffic models have been formulated [11,12,13,14]. Rules of these models comprise two steps: first, additional step takes into account lanes changing behaviour and second, basic step describes forward movement of a vehicle.

Cellular automata have also been used for junctions modelling. The simplest cellular model of a crossroad [15] did not take into account region inside the junction as well as priority rules. Vehicles were just randomly selected to pass the crossroad. In [16] other simple model has been proposed of cellular automaton with closed boundary conditions (ring of cells) for crossroads simulation. All junctions in this case were modelled as roundabouts. For more realistic simulations sophisticated models have been applied that include definitions of traffic regulations (priority rules, signs, signalisation) and allow for determination of actual junction capacity [17].

Little research work has been done in the application of microscopic cellular models for traffic control in road networks. Certain methods have been proposed of optimal route selection in urban networks [18]. The cellular automata model of road network has been used in this approach to evaluate current traffic conditions for particular connections. In [19] similar model was adopted to calculate basic parameters of coordination plan for signalised intersections network. However, the analysed cases were significantly simplified and far from practical solutions.

In the field of road traffic modelling several methods are known using cells defined as road segments for macroscopic flow description. Although the term cell is used in these methods, they are not derived from cellular automata theory. One cell in this case can be occupied by many vehicles thus its state is described using parameters of traffic stream (density, intensity). A model of this type was implemented for traffic control purposes in the UTOPIA method [2], different models have been introduced for highway traffic analysis [20].

## 3 Fuzzy cellular model of road traffic flow

The fuzzy cellular model of road traffic flow is proposed as an extension of the NaSch traffic model. It assumes a division of traffic lane into cells that correspond to road segments of equal length. The traffic state is described in discrete time steps. Vehicle position, its velocity and other parameters are modelled as fuzzy numbers defined on the set of integers.

State of a cell $c$ in time step $t$ is defined by fuzzy set of vehicles ($n$) that currently occupy this cell:

$$S_{c,t} = \left\{ \mu_{S_{c,t}}(n)/n \right\}. \tag{1}$$

Thus, one cell can be occupied by more than one vehicle in the same time. Conventionally, $\mu_A(x)$ denotes value of membership function of fuzzy set $A$ for an element $x$. Position of vehicle $n$ in time step $t$ is a fuzzy number defined on

the set of cells indexes ($c$):

$$P_{n,t} = \{\mu_{P_{n,t}}(c)/c\}. \tag{2}$$

Vehicle $n$ is described by its class and velocity $V_{n,t}$ (in cells per time step). The class determines properties of vehicle $n$: length $L_n$ (in cells), maximal velocity $V_n^{max}$, and acceleration $A_n$. All these quantities are expressed by fuzzy numbers. Velocity of the vehicle $n$ in time step $t$ is computed according to formula:

$$V_{n,t} = \widetilde{\min} \left\{ V_{n,t-1} \widetilde{+} A_n, G_{n,t}, V_n^{max} \right\}. \tag{3}$$

The tilde ($\sim$) symbol is used to distinguish operations on fuzzy numbers [21]. Gap $G_{n,t}$ is the number of free cells in front of vehicle $n$:

$$G_{n,t} = \widetilde{\min_{m \neq n}} \left\{ P_{m,t} \widetilde{-} L_m \widetilde{-} P_{n,t} : P_{m,t} \widetilde{>} P_{n,t} \right\}. \tag{4}$$

If there is no vehicle $m$ fulfilling the condition in (4), gap $G_{n,t}$ is assumed to be equal to the maximal velocity $V_n^{max}$.

Position of vehicle $n$ in the next time step ($t+1$) is computed on the basis of the model state in time $t$:

$$P_{n,t+1} = \widetilde{dil} \left( P_{n,t} \widetilde{+} V_{n,t} \right), \tag{5}$$

$\widetilde{dil}$ denotes fuzzy set dilation:

$$\mu_{\widetilde{dil}\left(P_{n,t} \widetilde{+} V_{n,t}\right)}(x) = \left[ \mu_{\left(P_{n,t} \widetilde{+} V_{n,t}\right)}(x) \right]^e, \tag{6}$$

where $0 < e \leq 1$.

Dilating the fuzzy set increases the fuzziness (uncertainty) of the vehicles position. This operation corresponds to the randomization step of traffic models based on NaSch cellular automaton. In the models that use s2s rules the randomization level decreases with increasing velocity of a vehicle as the random driver behaviours are more intense at low velocity range. To achieve similar effect for the presented model the exponent $e$ in (6) was defined as an increasing function of velocity. It was also assumed that when the maximal velocity is reached the vehicle position is no further dilated ($e = 1$). A simple linear dependency was used to control the dilation:

$$e = \alpha + \frac{1-\alpha}{\hat{v}_n^{max}} \hat{v}_{n,t}, \tag{7}$$

where $0 \leq \alpha \leq 1$ and $\hat{v}$ denotes defuzzified (crisp) value of velocity:

$$\hat{v}_{n,t} = \arg\max_y \mu_{V_{n,t}}(y). \tag{8}$$

Fig. 1 presents results of a traffic simulation that was performed using the fuzzy cellular model. Simulation was started with single vehicle ($n = 0$) stopped in the first cell ($c = 0$): $P_{0,0} = \{1/0\}$, $V_{0,0} = \{1/0\}$, the vehicle properties was set: $L_0 = \{1/0\}$, $V_0^{max} = \{0, 2/4; 1/5; 0, 2/6\}$, $A_0 = \{0, 2/0; 1/1; 0, 2/2\}$. Space-time diagrams in fig. 1 depict how the vehicle accelerates, its fuzzy positions are showed using gray levels. If the colour is darker for a cell, the value of membership function of fuzzy set $P_{0,t}$ is higher in this cell (white colour indicates empty cells, black indicates cells where $\mu_{P_{0,t}}(c) = 1$).

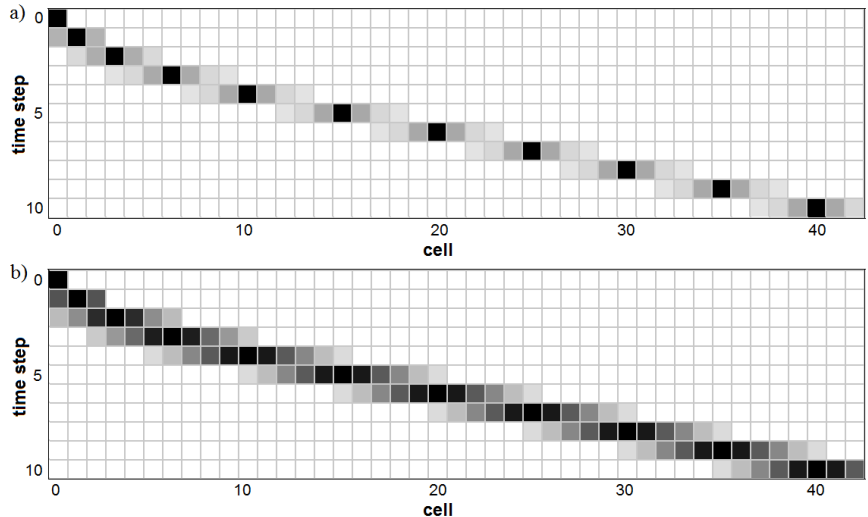

**Fig. 1.** Space-time diagram for the single-vehicle simulation: a) $\alpha = 0, 9$ b) $\alpha = 0, 1$

Results of the single vehicle movement simulation are compared in fig. 1 for two values of parameter $\alpha$, which was used for controlling the dilation (eq. 7). It shows that decreasing value of $\alpha$ causes increase in fuzziness of the vehicle position (higher level of model uncertainty).

## 4 Vehicles queue modelling

The fuzzy cellular model presented in previous section was applied for a simulation of vehicles queue discharge process that corresponds to real-life situations observed at approaches of signalised crossroads when green signal is given. During this experiment length of vehicles queue was analysed in sequence of time steps. This section includes discussion of the simulation results as well as their comparison with experimental data obtained using NaSch traffic model.

Only one class of vehicles was used in this simulation defining their maximal velocity: $V_{max} = \{0, 2/2; 1/3; 0, 2/4\}$ and remaining properties that were set identically to those of single-vehicle experiment reported in section 2. Dilation operation (6) was applied with parameter $\alpha = 0,9$. In the first step of simulation all fifty vehicles ($n = 0 \ldots 49$) were stopped in a queue: $P_{n,0} = \{1/n\}$, $V_{n,0} = \{1/0\}$, it means that at the beginning length of the queue was equal to the number of vehicles.

In subsequent time steps the traffic model was updated according to equations (3), (5) and the queue length $Q_t$ was evaluated. For the introduced traffic model $Q_t$ is defined as fuzzy value using following fuzzy rule:

if $veh\_0$ is $in\_queue$ and $veh\_1$ is $in\_queue$ and...and $veh\_x - 1$ is $in\_queue$

and $veh\_x$ is not $in\_queue$ and...and $veh\_m$ is not $in\_queue$ then $Q_t$ is $x$, (9)

where $veh\_n$ stands for "vehicle $n$", $m$ denotes number of vehicles and variable $in\_queue$ is determined by another fuzzy rule taking into account position and velocity of a vehicle:

$$\text{if } P_{n,t} \text{ is } n \text{ and } V_{n,t} \text{ is } 0 \text{ then } veh\_n \text{ is } in\_queue. \tag{10}$$

Results of vehicles queue length computations based on fuzzy cellular simulation are presented in fig 2 a), gray scale was used in this case to depict membership function value of $Q_t$ (darker colour correspond to higher value). These results are compared with experimental data on vehicles queue discharge that was collected from traffic simulation driven by NaSch cellular automaton (fig. 2 b).

Vehicles characteristics as well as starting conditions for the simulation using NaSch model were similar to those defined for fuzzy cellular simulation. Probabilistic parameter $p$ of the NaSch model was set to 0,2 and $v_{max}$ was 3. Simulation was executed over two hundred times to gather the data presented in fig. 2 b). Gray levels in this chart correspond to experimental probability of specific queue length evaluated for a given time step of the simulation.

The fuzzy cellular model was validated with regard to the traffic flow theory by comparing fundamental diagrams (dependency between traffic density and flow volume). Experimental data for the NaSch model (fig. 4 d) was collected from 200 executions of traffic simulation. Black dots in this diagram indicate traffic states that have experimental probability higher or equal 0,1, gray dots correspond with lower probabilities. Figure 4 c) includes diagram obtained for the fuzzy cellular model. Flow volume was determined as a fuzzy number for each value of traffic density. The black dots in this diagram indicate maximal values of the flow membership functions and the gray lines correspond to alfa-cuts computed using threshold value of 0,99.

Comparison of the simulation results (fig. 2) shows that proposed fuzzy cellular model adequately describes the process of traffic flow. It should be noted that single simulation using fuzzy cellular model gives comparable results to the distribution of experimental probability computed for many NaSch simulations. The proposed model inherently describes uncertainty of traffic states. Conventional cellular automaton needs many instances of the model and additional statistics to explore uncertainty of traffic parameters i.e. range of their possible values.

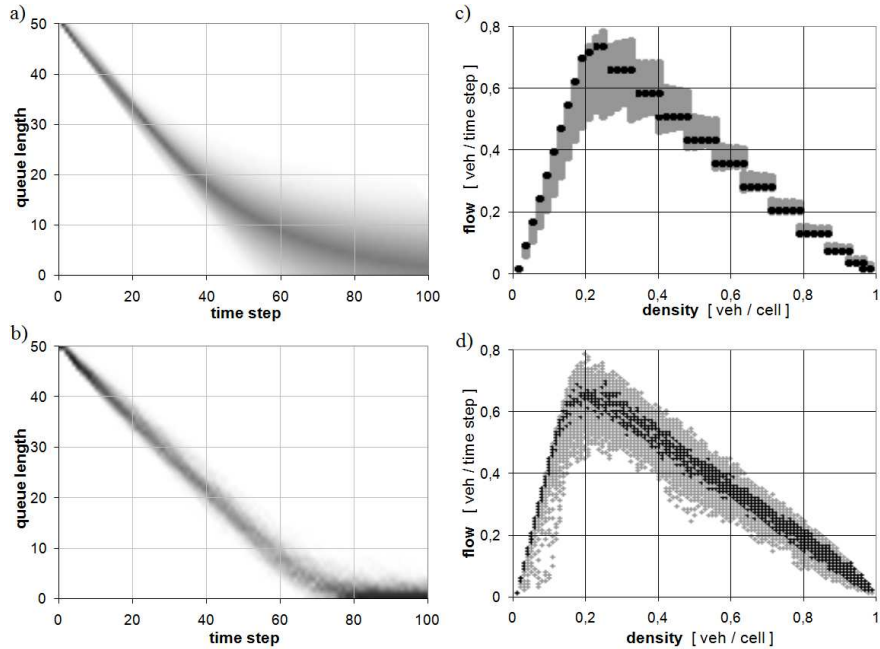

**Fig. 2.** Results of the vehicles queue discharge simulation

## 5 Conclusions

Fuzzy cellular model of road traffic was formulated on the basis of cellular automata approach. Parameters of vehicles are individually described by means of fuzzy numbers, thus various classes of vehicles can be modelled. Application of fuzzy sets theory allows to represent uncertainty of traffic states using single instance of the model. It takes into account random nature of traffic processes and makes the model suitable for collecting imprecise data from traffic detectors.

The uncertainty is described applying fuzzy definitions of vehicles parameters. E.g. uncertain position of a vehicle is expressed by a fuzzy number defined on the set of cells indexes. This uncertain position in the model can correspond with imprecise results of vehicle detection. Thus, the main advantage of fuzzy approach is the model's ability to utilise imprecise traffic data for on-line simulation.

Experimental results of the vehicles queue discharge simulation reported in this contribution show that fuzzy cellular model adequately describes the process of traffic flow. The queue lengths and fundamental diagrams were analysed in this experiment and compared to the results of NaSch cellular automata model. Further tests are necessary to evaluate the model applicability for other real-traffic situations. Planned research will also involve design of communication procedures that are needed to input data from traffic detectors into the model.

# References

1. Martin, P.T., Hockaday, S.L.M.: SCOOT: An update. ITE Journal, vol. 65, no. 1, pp. 44–48 (1995)
2. Mauro, V., Taranto, C.: UTOPIA. In: Proceedings of the 6th IFAC/IFORS Conf. on Control, Computers and Communications in Transport, pp. 245–252. Paris (1989)
3. Płaczek, B., Staniek, M.: Model Based Vehicle Extraction and Tracking for Road Traffic Control. In: Kurzyński M. et al. (eds.) Advances in Soft Computing. Computer Recognition Systems 2, pp. 844–851. Springer-Verlag, Berlin Heidelberg (2007)
4. Nagel, K., Schreckenberg, M.: A cellular automaton model for freeway traffic. J. Physique I 2, pp. 2221–2241 (1992)
5. Płaczek, B.: The method of data entering into cellular traffic model for on-line simulation. In: Piecha, J. (ed.) Trans. on Transport Systems Telematics, pp. 34–41. Publishing House of Slesian Univ. of Technology, Gliwice (2006)
6. Barlovic, R., Santen, L., Schadschneider, A., Schreckenberg, M.: Metastable states in cellular automata for traffic flow. The European Physical Journal B, vol. 5, issue 3, pp. 793–800 (1998)
7. Chowdhury, D., Santen, L., Schadschneider, A.: Statistical physics of vehicular traffic and some related systems. Physic Reports, vol. 329, pp. 199–329 (2000)
8. Emmerich, H., Rank, E.: An improved cellular automaton model for traffic flow simulation. Physica A, vol. 234, no. 3-4, pp. 676–686 (1997)
9. Pottmeier, A., Berlovic, R., Knopse, W., Schadschneider, A., Schreckenberg, M.: Localized defects in a cellular automaton model for traffic flow with phase separation. Physica A, vol. 308, no. 1-4, pp. 471–482 (2002)
10. Shih-Ching, L., Chia-Hung, H.: Cellular Automata Simulation for Traffic Flow with Advanced Control Vehicles. In: The 11th IEEE Int. Conf. on Computational Science and Engineering Workshops, pp. 328–333. IEEE (2008)
11. Rickert, M., Nagel, K., Schreckenberg, M., Latour, A.: Two Lane Traffic Simulations using Cellular Automata. Physica A, vol. 231, no. 4, pp. 534–550 (1995)
12. Knopse, W., Santen L., Schadschneider, A., Schreckenberg, M.: Disorder in cellular automata for two-lane traffic. Physica A, vol. 265, issue 3-4, pp. 614–633 (1999)
13. Wagner, P., Nagel, K., Wolf, D.E.: Realistic Multi-Lane Traffic Rules for Cellular Automata. Physica A, vol. 234, no. 3-4, pp. 687–698 (1996)
14. Xianchuang, S., Xiaogang, J., Yong M., Bo P.: Study on Asymmetric Two-Lane Traffic Model Based on Cellular Automata. In: Sunderam, V.S. et al. (eds.): ICCS 2005. LNCS, vol. 3514, pp. 599–606. Springer, Heidelberg (2005)
15. Rickert, M., Nagel, K.: Experiences with a Simplified Microsimulation for the Dallas/Fort Worth Area. Int. J. of Modern Physics C, vol. 8, no. 3, pp. 483-503 (1997)
16. Dupuis, A., Chopard, B.: Parallel simulation of traffic in Geneva using cellular automata. In: Kuhn, E. (ed.) Virtual shared memory for distributed architectures, pp. 89–107. Nova Science Publishers , New York (2001)
17. Esser, J., Schreckenberg, M.: Microscopic simulation of urban traffic based on cellular automata. Int. J. of Modern Physics C, vol. 8, no. 5, pp. 1025–1036 (1997)
18. Wahle, J., Annen, O., Schuster, Ch., Neubert, L., Schreckenberg, M.: A dynamic route guidance system based on real traffic data. European Journal of Operational Research, vol. 131, no. 2, pp. 302–308 (2001)
19. Brockfeld, E., Barlovic, R., Schadschneider, A., Schreckenberg, M.: Optimizing traffic lights in a cellular automaton model for city traffic. Physical Review E, vol. 64, 056132 pp. 1–12 (2001)

20. Daganzo, C.: The cell transmission model. Part II: Network traffic. Transportation Research B, vol. 29, no. 2, pp. 79–93 (1995)
21. Dubois, D., Prade, H.: Operations on fuzzy numbers. International Journal of Systems Science, vol. 9, no. 6, pp. 613–626 (1978)